\documentclass[11pt]{iopart}

\usepackage{graphicx}
\usepackage{latexsym}
\usepackage{amssymb}
\usepackage{amsfonts}
\usepackage{color}
\usepackage[utf8]{inputenc}

\newcommand{\ket}[1]{\vert #1 \rangle}
\newcommand{\bra} [1] {\langle #1 \vert}
\newcommand{\proj}[1]{\ket{#1}\bra{#1}}
\newcommand{\kb}[2]{\left| #1 \vphantom{#2} \right>\left< #2 \vphantom{#1} \right|} % for Dirac ketbras

\renewcommand{\bs}{U^{\mathrm{BS}}_{\eta}}
\newcommand{\bsd}{U^{\mathrm{BS} \dagger}_{\eta}}
\newcommand{\tms}{U^{\mathrm{TMS}}_{\lambda}}
\newcommand{\tmsd}{U^{\mathrm{TMS} \dagger}_{\lambda}}

\newtheorem{theo}{Theorem}

\newtheorem{corollary}{Corollary}

\begin{document}

%%%%%%%%%%%%%%%%%%%%%%%%%%%%%%%%%%%%%%%%% Title and authors

\title{Fock majorization in bosonic quantum channels with a passive environment}

\author{Michael G. Jabbour and Nicolas J. Cerf}
\address{Quantum Information and Communication, \'Ecole polytechnique de Bruxelles, CP~165, Universit\'e libre de Bruxelles, 1050 Bruxelles, Belgium}
\ead{mjabbour@ulb.ac.be}
\vspace{10pt}

%\begin{indented}
%\item[]April 2017
%\end{indented}

%%%%%%%%%%%%%%%%%%%%%%%%%%%%%%%%%%%%%%%%% Abstract

\begin{abstract}
We introduce a class of quantum channels called passive-environment bosonic channels. These channels are relevant from a quantum thermodynamical viewpoint because they correspond to the energy-preserving linear coupling of a bosonic system with a bosonic environment that is in a passive state (no energy can be extracted from it by using a unitary transformation) followed by discarding the environment. The Fock-majorization relation defined in [New J. Phys. \textbf{18}, 073047 (2016)] happens to be especially useful in this context as, unlike regular majorization, it connects the disorder of a state together with its energy. Our main result here is the preservation of Fock majorization across all passive-environment bosonic channels. This implies a similar preservation property for regular majorization over the set of passive states, and it also extends to passive-environment channels whose Stinespring dilation involves an active Gaussian unitary. Beyond bosonic systems, the introduced class of passive-environment operations naturally generalizes thermal operations and is expected to provide new insights into the thermodynamics of quantum systems.
\end{abstract}

%%%%%%%%%%%%%%%%%%%%%%%%%%%%%%%%%%%%%%%%% Body

\section{Introduction}

Quantum thermodynamics has become a very active research area over the last years, aiming at a better understanding of thermal operations on individual quantum systems at the microscopic scale, see e.~g.  \cite{qthermo1,qthermo2,qthermo3,qthermo4}. Among the objectives that are pursued, finding conditions to discriminate the allowed operations from the forbidden ones is of key importance, with a milestone in this direction being the recently uncovered existence of several second laws of thermodynamics \cite{qthermo5}. In this context, majorization theory \cite{Ineq} has proven to be a powerful tool as it allows one to compare states in terms of disorder, which is a primordial concept when studying thermal operations on quantum systems (see, e.~g., the notion of thermomajorization \cite{qthermo2}). Although most works in quantum thermodynamics have considered discrete (often finite-dimensional) quantum systems, we turn here to continuous-variable bosonic (infinite-dimensional) quantum systems.

The use of majorization relations to express conditions on the interconvertibility between quantum entangled states  \cite{NielsenLOCC}
has successfully been extended to probe the interconvertibility between Gaussian bosonic entangled states \cite{Michael2015}. The properties of Gaussian bosonic channels have also been characterized using majorization theory \cite{Michael2016}. Here, we go a step further and exploit the tools based on majorization for bosonic channels in a thermodynamical context. The evolution of a quantum thermodynamical system can indeed be viewed as a completely-positive trace-preserving map applied to the system, that is, a quantum channel. We focus in this paper on bosonic quantum channels that are Gaussian-dilatable (i.e., a Gaussian unitary can be used in the Stinespring dilation of the channel) and involve a \textit{passive} environment (i.e., no energy can be extracted by applying a unitary on the environment state).

We address the question of whether a majorization relation is transferred across such a bosonic channel, viewed as a thermodynamical operation. Our results rely on the notion of Fock-majorization (or energy majorization) \cite{Michael2016}, and imply that any two input states that obey a Fock-majorization relation are transformed into output states obeying the same relation. This property thus holds for a large class of thermodynamically relevant channels, going beyond the special case of Gaussian bosonic channels that was investigated in \cite{Michael2016} (passive-environment channels are non-Gaussian channels).  

In Section 2, we summarize the notion of passive states and their role in quantum thermodynamics. We then define the class of bosonic quantum channels with a passive environment, which is a natural generalization of the noisy operations and thermal operations used for modelling the dynamics of quantum thermodynamical systems. In Section 3, we review the Fock-majorization relation defined in \cite{Michael2016} and prove two necessary and sufficient conditions that are equivalent to the original formulation of Fock majorization, thereby making a close parallel with the theory of regular majorization. In Section 4, we prove the preservation of Fock-majorization relations across bosonic channels characterized by a passive Gaussian unitary and a passive environment, and then discuss the implication for regular majorization preservation over the set of passive states. A main ingredient of our proofs derives from the analysis of the generating function of the matrix elements of Gaussian unitaries in the Fock basis, which yields useful recurrence equations on these non-Gaussian objects \cite{RecBSTMS}. In Section 5, we extend these results to the passive-environment channels where the Gaussian unitary is active. Finally, in Section 6, we give our conclusions.

\section{Passive states and passive-environment bosonic channels}

Passive states are interesting when studying quantum systems from a thermodynamical point of view. They are defined as those quantum states from which no work can be extracted under Hamiltonian processes, making them the most stable states among all states that are reachable through a unitary transformation \cite{Passive1}. As a result, a passive state, denoted as $\rho^{\downarrow}$, is diagonal in the eigenbasis of the Hamiltonian of the system and is characterized by non-increasing eigenvalues when the energy of the corresponding eigenvectors increases. Mathematically speaking, it can be written as 
\begin{equation}
\rho^{\downarrow} = \sum_i \lambda^{\downarrow}_i \, \proj{e_i}  \mathrm{~~~~~~with~} \lambda^{\downarrow}_{i+1} \leq \lambda^{\downarrow}_{i} \mathrm{~if~}e_{i+1} > e_{i} ,
\end{equation}
where $\ket{e_i}$ are the eigenvectors and $e_i$ the corresponding eigenvalues of the Hamiltonian of the system. Interestingly, one can often  ``activate'' the work extraction from a passive state by jointly acting on it and an ancillary system with a joint unitary \cite{Passive1,Passive2}. Suppose one has access to $n$ replicas of the passive state (in this example, the ancilla consists of $n-1$ replicas), then the joint system may not be passive anymore, allowing one to extract work from the joint system. For a sufficiently large $n$, this is actually the case for almost all passive states except for thermal states. The latter are a special case of passive states whose eigenspectrum is given by a geometric distribution, which is characterized by a single parameter (e.g., the temperature). Remarkably, the tensor product of $n$ replicas of a thermal state remains passive and no work can be extracted from it. Furthermore, if one fixes the von Neumann entropy of a state, the thermal state happens to be the state with the lowest energy among all states (including the passive states) having this entropy. Hence, we may categorize thermal states as the most stable states among all passive states having the same entropy.

Passive states also arise in the context of modelling the dynamics of quantum thermodynamical systems, where some specific passive states are usually chosen as free ``resources". When constructing a resource theory, one needs to define the set of allowed (free) state transformations \cite{Thermo}. This can be done by combining the following operations: composing the state with a fixed environment (viewed as a bath), acting on the resulting joint state with a unitary (which is usually chosen to conserve the energy), and finally discarding the environment. The environment is usually chosen to be thermal, which is a reasonable physical assumption. Still, one can also construct a simpler, less realistic model by choosing the maximally mixed state for the environment. By doing so, one obtains so-called \textit{noisy operations} (NO), which have the form
\begin{equation}
\mathcal{C}_{\mathrm{NO}}(\rho_{\mathrm{S}}) = \mathrm{Tr}_{\mathrm{E}} \left[ U_{\mathrm{SE}} \left( \rho_{\mathrm{S}} \otimes \frac{\mathbb{I}_{\mathrm{E}}}{n_{\mathrm{E}}} \right) U_{\mathrm{SE}}^{\dagger} \right],
\label{NO}
\end{equation}
where $\rho_{\mathrm{S}}$ is the state of the system, $\mathbb{I}_{\mathrm{E}}$ is the identity defined on the environment of dimension $n_{\mathrm{E}}$, and $U_{\mathrm{SE}}$ is an energy-conserving unitary acting on the system and the environment. When a state is transformed according to $\mathcal{C}_{\mathrm{NO}}$, the input can be shown to majorize the output for large enough $n_E$ \cite{MajNO}. This can be intuitively understood by noticing that a state undergoing such a transformation gets more mixed. A more realistic model is obtained by choosing a thermal state $\tau_{\mathrm{E}}$ as an environment, resulting into the so-called \textit{thermal operations} (TO),
\begin{equation}
\mathcal{C}_{\mathrm{TO}}(\rho_{\mathrm{S}}) = \mathrm{Tr}_{\mathrm{E}} \left[ U_{\mathrm{SE}} \left( \rho_{\mathrm{S}} \otimes \tau_{\mathrm{E}} \right) U_{\mathrm{SE}}^{\dagger} \right].
\label{TO}
\end{equation}
A similar input-output relation can be proven in the case of thermal operations $\mathcal{C}_{\mathrm{TO}}$, with majorization being replaced by thermomajorization  \cite{qthermo2}.  Roughly speaking, this corresponds to majorization after a rescaling of the eigenvalues of the state using those of the thermal environment.

In this paper, we introduce a class of quantum channels that generalizes Eqs. (\ref{NO}) and (\ref{TO}), where the environment is chosen to be in any passive state (note that in $\mathcal{C}_{\mathrm{NO}}$ and $\mathcal{C}_{\mathrm{TO}}$, the environment is in a special case of a passive state).   Since we focus on bosonic systems, we choose the environment  to be passive in the eigenbasis of the Hamiltonian of the harmonic oscillator (i.e., the Fock basis), and fix the unitary $U_{\mathrm{SE}}$ to be a beam splitter (i.e., the realization of an energy-conserving linear coupling between bosonic systems). The result is a thermodynamical operation that we call a passive-environment bosonic channel, which is of the form
\begin{equation}
\mathcal{B}^{\downarrow}_{\eta}(\rho_{\mathrm{S}}) = \mathrm{Tr}_{\mathrm{E}} \left[ \bs \left( \rho_{\mathrm{S}} \otimes \sigma^{\downarrow}_{\mathrm{E}} \right) \bsd \right],
\label{PO}
\end{equation}
where $\sigma^{\downarrow}_{\mathrm{E}} = \sum_i \lambda^{\downarrow}_i \, \proj{i}$ is the passive state of the environment, with $\lambda^{\downarrow}_{i+1} \leq \lambda^{\downarrow}_i$ and $\ket{i}$ denoting Fock states. The unitary $\bs$ corresponds to a beam splitter (BS) of transmittance $\eta$ (hence the symbol $\mathcal{B}$ for the channel). It couples the system mode $S$ with the passive environment mode $E$ (hence the arrow in the notation $\mathcal{B}^{\downarrow}$) through the relation $\hat{a}_\mathrm{S} \rightarrow \sqrt{\eta} \, \hat{a}_\mathrm{S} + \sqrt{1-\eta} \, \hat{a}_\mathrm{E}$,
where $\hat{a}_\mathrm{S}$ are $\hat{a}_\mathrm{E}$ are the bosonic mode operators for the system and  environment, respectively \cite{Weedbrook2011}.

Note that in contrast with $\mathcal{C}_{\mathrm{TO}}$, which corresponds to a Gaussian channel in the case of bosonic systems, the map $\mathcal{B}^{\downarrow}_{\eta}$ effects a non-Gaussian channel since the system is coupled (via a Gaussian unitary) to an environment state that is generally non-Gaussian (this is called a Gaussian-dilatable channel since there exists a Stinespring dilation of the channel admitting a Gaussian unitary). In this sense, our study of majorization relations for $\mathcal{B}^{\downarrow}_{\eta}$ in Section 4 generalizes the earlier study for Gaussian channels \cite{Michael2016}. To be more general, we will also consider in Section 5 the class of Gaussian-dilatable channels with an active Gaussian unitary,  namely a  two-mode squeezer (TMS), the environment being again passive. These maps are noted $\mathcal{A}^{\downarrow}_{G}$, where $G$ is the gain of the two-mode squeezer and the corresponding unitary $U^{\mathrm{TMS}}_{G}$ couples the system mode $S$ with the passive environment mode $E$ through the relation $\hat{a}_\mathrm{S} \rightarrow \sqrt{G} \, \hat{a}_\mathrm{S} + \sqrt{G-1} \, \hat{a}^{\dagger}_\mathrm{E}$. It turns out that $\mathcal{A}^{\downarrow}_{G}$ exhibits similar properties to those of $\mathcal{B}^{\downarrow}_{\eta}$ in terms of majorization.

\section{Fock-majorization relation}

Before turning to the Fock-majorization relation, it is adequate to recall a few basics on the theory of majorization applied to quantum systems. Majorization provides a pre-order relation on quantum states, allowing us to compare them in terms of disorder. We say that a state $\rho$ majorizes another state $\sigma$, denoted as $\rho \succ \sigma$, when
\begin{equation}
\sum_{i=1}^n r^{\downarrow}_i \geq \sum_{i=1}^n s^{\downarrow}_i, \quad \forall \ n \geq 1,
\label{definition-majorization}
\end{equation}
where $\bi{r}^{\downarrow}$ ($\bi{s}^{\downarrow}$) is the vector of eigenvalues of $\rho$ ($\sigma$) arranged in non-increasing order. Whenever Eq.~(\ref{definition-majorization}) is verified, it means that $\bi{s} = \rm{D} \, \bi{r}$, where  $\bi{r}$ ($\bi{s}$) is the vector of eigenvalues of $\rho$ ($\sigma$) and D is a bistochastic matrix, so that state $\sigma$ can be obtained from state $\rho$ by applying a random mixture of unitaries ($\sigma$ is more disordered than $\rho$). In addition, $\rho \succ \sigma$ is also equivalent to ${\rm Tr} f(\rho) \geq {\rm Tr} f(\sigma) $ for any convex function $f: \mathbb{R}\to \mathbb{R}$, which introduces a structure in terms of convex functions.

The concept of Fock-majorization  was introduced in \cite{Michael2016}, and can more generally be viewed as energy-majorization when the Hamiltonian is not the one of the harmonic oscillator. For a harmonic oscillator (or a bosonic mode), we say that a state $\rho$ Fock-majorizes another state $\sigma$, denoted as $\rho \succ_{\rm{F}} \sigma$, when
\begin{equation}
\rm{Tr}(P_n \, \rho) \geq \rm{Tr}(P_n \, \sigma), \quad \forall \ n \geq 0,
\end{equation}
where $P_n = \sum_{i=0}^n \proj{i}$ is a projector onto the space spanned by the $n+1$ first Fock states $\ket{i}$ (which are the eigenstates of the Hamiltonian of the harmonic oscillator). This (pre)order relation only depends on the diagonal elements of $\rho$ and $\sigma$ in the eigenbasis of the Hamiltonnian, i.e., the Fock basis. In contrast with regular majorization, these diagonal elements are not ordered by decreasing values, but instead by increasing photon number\footnote{In general, such a definition of energy-majorization without prior sorting makes sense because there exists a natural way of ordering the elements, here the energy.}. As mentioned in \cite{Michael2016}, Fock-majorization bears some similarity to the relation called ``upper-triangular majorization" introduced in \cite{qt1}. There, it was shown that two states obeying such a relation can be related by a so-called ``cooling" map, which happens to be a special case of the thermal operations (\ref{TO}) when the environment is set to zero temperature (it is in the vacuum state). Instead, we show that Fock-majorization can be interpreted as a relation indicating the existence of a ``heating" (or ``amplifying") map between the two states, corresponding to a ``lower-triangular majorization'', as exhibited by the following theorem.

\begin{theo} \label{theoFM2}
Two states $\rho$ and $\sigma$ whose diagonal elements in the Fock basis are given by the respective vectors $\bi{r}$ and $\bi{s}$ obey $\rho \succ_{\rm{F}} \sigma$ if and only if there exists a column-stochastic lower-triangular matrix $\bi{L}$ such that $\bi{s} = \bi{L} \bi{r}$, with $L_{ij} \ge 0, \forall \ i \ge j \ge 1$, and $\sum_{i=j}^d L_{ij} = 1, \forall \ j \ge 1$.
\end{theo}

Note that the indices range from 1 to $d$, corresponding to Fock states ranging from $\ket{0}$ to $\ket{d-1}$. At the end of the proof, we must take the limit $d\to \infty$ resulting in the full Fock space. Interestingly, Theorem \ref{theoFM2} reminds us of the property that two probability distributions related by a majorization relation can be connected through a bistochastic matrix (here, it is replaced by a column-stochastic lower-triangular matrix).

\noindent \textit{Proof}. The proof we give here is slightly simpler than the corresponding one given in \cite{qt1} for the ``cooling map". First, suppose there exists a matrix $\bi{L}$ satisfying the conditions of Theorem \ref{theoFM2}. In this case, we have
\begin{equation}
\sum_{i=1}^m s_i = \sum_{i=1}^m \sum_{j=1}^i L_{ij} r_j = \sum_{j=1}^m r_j \sum_{i=j}^m L_{ij}, \quad \quad \forall \ m \ge 1.
\end{equation}
Since $\sum_{i=1}^d L_{ij} = 1, \forall \ j \ge 1$, we have that $\sum_{i=1}^m L_{ij} \le 1, \forall \ j \ge 1$ and $\forall \ m \ge 1$ (with the condition that $L_{ij} \ge 0, \forall \ i \ge j \ge 1$). This yields the relation $\sum_{i=1}^m s_i \le \sum_{j=1}^m r_j, \forall \ m \ge 1$, which concludes the first part of the proof.

Now, suppose that $\rho \succ_{\rm{F}} \sigma$. We are going to construct $\bi{s}$ step by step starting from $\bi{r}$, using a succession of lower-triangular matrices. Starting with the vector $\bi{r} = \left(r_1, r_2, \cdots r_d\right)^{\rm T}$, we first define $\bi{w}^{(1)} = \left(s_1, (r_2 + r_1 - s_1), r_3, \cdots r_d\right)^{\rm T}$. Since $\bi{r} \succ_{\rm{F}} \bi{s}$, we have that $r_2 + r_1 - s_1 \ge s_2 \ge 0$, which means that $\bi{w}^{(1)}$ is a well-defined vector of probability distribution, its elements being non-negative and summing to one. Similarly, we construct $\bi{w}^{(2)} = \left(s_1, s_2, (r_3 + r_2 + r_1 - s_1 - s_2), r_4, \cdots r_d\right)^{\rm T}$, which also represents a well-defined probability distribution for the same reasons. More generally, we define $\bi{w}^{(k)} = \left(s_1, s_2, \cdots s_k, (\sum_{j=1}^{k+1} r_j - \sum_{j=1}^k s_j), r_{k+2}, \cdots r_d\right)^{\rm T}$, each of the $\bi{w}^{(k)}$ representing a well-defined probability distribution, for $k \le d$. Furthermore, we end up with $\bi{w}^{(d)} = \bi{s}$, which we wanted to reach starting from $\bi{r}$. Now, we show that each $\bi{w}^{(k)}$ is related to the corresponding $\bi{w}^{(k-1)}$ through a lower-triangular matrix, which has all its diagonal elements equal to one, apart from the one on column $k$. In order to do this, write
\[
    \left\{
    \begin{array}{c l}	
         w^{(k)}_k & = \mu_1 w^{(k-1)}_k \\
         w^{(k)}_{k+1} & = \mu_2 w^{(k-1)}_k + \mu_3 w^{(k-1)}_{k+1}
    \end{array}\right.
\]
which correspond to
\[
    \left\{
    \begin{array}{c l}	
         s_k & = \mu_1 \left( \sum_{j=1}^{k} r_j - \sum_{j=1}^{k-1} s_j \right) \\
         \sum_{j=1}^{k+1} r_j - \sum_{j=1}^k s_j & = \mu_2 \left( \sum_{j=1}^{k} r_j - \sum_{j=1}^{k-1} s_j \right) + \mu_3 r_{k+1}
    \end{array}\right.
\]
If we want the matrix which relates $\bi{w}^{(k-1)}$ to $\bi{w}^{(k)}$ to be column-stochastic (as well as lower-triangular), we need $\mu_3 = 1$. This is also consistent with the fact that the diagonal element of column $k+1$ should be equal to one, as we chose earlier. We still need to check if both our equations are compatible with the fact that $\mu_1 \ge 0$, $\mu_2 \ge 0$, and $\mu_1 + \mu_2 = 1$. According to our first equation,
\begin{equation}
\mu_1 = \frac{s_k}{\sum_{j=1}^{k} r_j - \sum_{j=1}^{k-1} s_j}.
\end{equation}
Since $\sum_{j=1}^{k} r_j - \sum_{j=1}^{k-1} s_j \geq s_k$, we indeed have that $\mu_1$ is non-negative and smaller than one. The second equation tells us that
\begin{equation}
\mu_2 = \frac{\sum_{j=1}^{k+1} r_j - \sum_{j=1}^k s_j - r_{k+1}}{\sum_{j=1}^{k} r_j - \sum_{j=1}^{k-1} s_j} = \frac{\sum_{j=1}^{k} r_j - \sum_{j=1}^k s_j}{\sum_{j=1}^{k} r_j - \sum_{j=1}^{k-1} s_j}
\end{equation}
which is non-negative and smaller than one for the same reasons. Now, it is also trivial to see that $\mu_1 + \mu_2 = 1$, which means that the matrix relating $\bi{w}^{(k-1)}$ and $\bi{w}^{(k)}$ has indeed non-negative elements, is column stochastic, and is lower-triangular. This also means that $\bi{r}$ can be related to $\bi{s}$ using a product of lower-triangular matrices, which is itself lower-triangular (and which is column-stochastic and has non-negative elements, as needed). Taking the limit $d\to \infty$  ends the proof. $\Box$

In \cite{Michael2016}, a connection between Fock-majorization and energy was exhibited. It was shown that if $\rho \succ_{\rm{F}} \sigma$, then the energy of $\sigma$ is greater than or equal to the one of $\rho$, i.e, $\rho \succ_{\rm{F}} \sigma \Rightarrow \rm{Tr}(H \rho) \leq \rm{Tr}(H \sigma)$, where $H$ is the Hamiltonian of the harmonic oscillator. Here, we go a step further by generalizing this property to functions of $H$ and turning it into an equivalence.

\begin{theo} \label{theoFM1}
Two states $\rho$ and $\sigma$ obey $\rho \succ_{\rm{F}} \sigma$ if and only if $\rm{Tr}[f(H) \, \rho] \leq \rm{Tr}[f(H) \, \sigma]$ for any function $f: \mathbb{R} \rightarrow \mathbb{R}$ which is continuous and increasing.
\end{theo}
Again, this property can be viewed as the counterpart of the equivalence between regular majorization and the condition in terms of convex functions.

\noindent \textit{Proof.} First, suppose $\rho \succ_{\rm{F}} \sigma$. Again, denote by $\bi{r}$ and $\bi{s}$ the vectors of diagonal elements of $\rho$ and $\sigma$ in the Fock-basis, and fix their dimension to be $d$ (at the end of the proof, we take the limit $d\to \infty$.) We need to show that, for any function $f: \mathbb{R} \rightarrow \mathbb{R}$ which is continuous and increasing,
\begin{equation}
\sum_{i=1}^d f(i) r_i - \sum_{i=1}^d f(i) s_i \le 0.
\end{equation}
According to Theorem \ref{theoFM2}, there exists a lower-triangular matrix $\bi{L}$ with non-negative elements, which is column-stochastic, and such that $\bi{s} = \bi{L} \bi{r}$. Thus,
\begin{equation}
\sum_{j=1}^d f(j) s_j = \sum_{j=1}^d f(j) \sum_{i=1}^j L_{ji} r_i = \sum_{i=1}^d r_i \sum_{j=i}^d f(j) L_{ji}
\end{equation}
meaning that
\begin{equation} \label{eqFM12}
\sum_{i=1}^d f(i) r_i - \sum_{i=1}^d f(i) s_i = \sum_{i=1}^d r_i \left[ f(i) - \sum_{j=i}^d f(j) L_{ji} \right].
\end{equation}
Now,
\begin{equation}\label{eqFM11}
f(i) - \sum_{j=i}^d f(j) L_{ji} = \sum_{j=i}^d L_{ji} f(i) - \sum_{j=i}^d f(j) L_{ji} = \sum_{j=i}^d L_{ji} \left[ f(i) - f(j) \right]
\end{equation}
Since $f$ is increasing, we have that $f(i) - f(j) \le 0$ when $j \ge i$. Furthermore, all the elements of $\bi{L}$ are non-negative, meaning that the left-hand side of Eq. (\ref{eqFM11}) is negative or equal to zero. Consequently, the left-hand side of Eq. (\ref{eqFM12}) is also negative or equal to zero. This concludes the first part of the proof.

Now, suppose that $\sum_{i=1}^d f(i) r_i \le \sum_{i=1}^d f(i) s_i$, for any function $f: \mathbb{R} \rightarrow \mathbb{R}$ which is continuous and increasing. Choose the series of functions $f_k: \mathbb{R} \rightarrow \mathbb{R}$ which verify
\[
    f_k(x)=\left\{
    \begin{array}{c l}	
         - 1 & \mathrm{if} \ x \le k, \\
         0 & \mathrm{else}.
    \end{array}\right.
\]
We can always find continuous and increasing functions which verify these properties. This means that
\begin{equation}
\sum_{i=1}^d f_k(i) r_i \le \sum_{i=1}^d f_k(i) s_i, \forall \ k \quad \Rightarrow \quad \sum_{i=1}^k r_i \ge \sum_{i=1}^k s_i, \forall \ k,
\end{equation}
which essentially means that $\rho \succ_{\rm{F}} \sigma$. This concludes the second part of the proof. $\Box$

\section{Fock-majorization preservation in passive-environment channels}

The notion of a majorization-preserving quantum channel was defined in \cite{Michael2016}. A channel $\Phi$ is called majorization-preserving whenever it is such that if $\rho \succ \sigma$, then $\Phi[\rho] \succ \Phi[\sigma]$. The central result of \cite{Michael2016} was that all  (phase-insensitive and phase-conjugate) Gaussian bosonic  channels $\Phi_G$ are majorization-preserving over the set of passive states. That is, given two passive states $\rho^{\downarrow}$ and  $\sigma^{\downarrow}$, if $\rho^{\downarrow} \succ \sigma^{\downarrow}$, then $\Phi_G[\rho^{\downarrow}] \succ \Phi_G[\sigma^{\downarrow}]$ for all $\Phi_G$. The proof relied on the Fock-majorization relation and the fact that it coincides with regular majorization for passive states (i.e, $\rho^{\downarrow} \succ \sigma^{\downarrow} \Leftrightarrow \rho^{\downarrow} \succ_\mathrm{F} \sigma^{\downarrow}$). As a matter of fact, Gaussian channels $\Phi_G$ were first proven to be Fock-majorization preserving, where 
a Fock-majorization preserving channel $\Phi$ is of course defined as a channel such that if $\rho \succ_{\rm{F}} \sigma$, then $\Phi[\rho] \succ_{\rm{F}} \Phi[\sigma]$. The preservation of  Fock-majorization across channels $\Phi_G$ was actually the key result of \cite{Michael2016}, from which the rest follows. It was proven based on the following theorem.
\begin{theo}[\cite{Michael2016}]
A channel $\Phi$ satisfying the condition $\bra{n} \, \Phi[\ket{i}\bra{j}] \, \ket{n}=0$,  $\forall i \ne j$, $\forall n$ is Fock-majorization preserving if and only if it obeys the ladder of Fock-majorization relations
\begin{equation}
\Phi\big[ \proj{i} \big]  \succ_\mathrm{F} \Phi\big[ \proj{i+1} \big], \quad  \forall i\ge 0.   \label{ladder}
\end{equation}
\label{theoMichael2016_2}
\end{theo}

In \cite{Michael2016}, all Gaussian  channels $\Phi_G$ were indeed shown to verify Eq.~(\ref{ladder}). Since they form a special case of passive-environment bosonic channels\footnote{In particular, the lossy Gaussian channels (i.e., channels whose Stinespring dilation gives a beam splitter) are passive-environment bosonic channels of the form (\ref{PO}), where the environment is chosen to be in a thermal (hence, passive) Gaussian state.}, it is therefore natural to investigate whether the Fock-majorization preservation property extends to all passive-environment bosonic channels $\mathcal{B}^{\downarrow}_{\eta}$  (and similarly $\mathcal{A}^{\downarrow}_{G}$).

In order to prove this, we again recourse to Theorem \ref{theoMichael2016_2}, with a minor caveat. Indeed, the proof of Theorem \ref{theoMichael2016_2} in \cite{Michael2016} did not mention the condition $\bra{n} \, \Phi[\ket{i}\bra{j}] \, \ket{n}=0$,  $\forall i \ne j$, $\forall n$  since only Fock-diagonal states were considered at the input of channel $\Phi_G$. However, Theorem \ref{theoMichael2016_2} also applies to input states that are non-diagonal in the Fock basis as long as the above condition is fulfilled (i.e., the non-diagonal elements of the input state do not contribute to the diagonal elements of the output state, which are the only ones that matter in the Fock-majorization relation). 
As shown in \ref{appNonDiag}, this condition is verified for Gaussian-dilatable channels with a passive environment, so before using Theorem \ref{theoMichael2016_2} for these channels we are left with having to prove the following theorem.

% $\Phi \equiv \mathcal{B}^{\downarrow}_{\eta}$

\begin{theo} Passive-environment bosonic channels $\mathcal{B}^{\downarrow}_{\eta}$ exhibit the ladder of Fock-majorization relations
\begin {equation}
\mathcal{B}^{\downarrow}_{\eta} \big[ \proj{i} \big]  \succ_\mathrm{F} \mathcal{B}^{\downarrow}_{\eta} \big[ \proj{i+1} \big] , \quad  \forall i\ge 0.
\label{EQlemPOLFM}
\end{equation}
\label{lemPOLFM}
\end{theo}
\textit{Proof.} We begin by proving the ladder of Fock-majorization relations for a passive channel $\mathcal{B}^{[K]}_{\eta}$ characterized by an environment that is a projector onto the space spanned by the $K+1$ first Fock states $\ket{k}$, i.e,
\begin{equation}
\mathcal{B}^{[K]}_{\eta}(\rho) = \mathrm{Tr}_E \left[ \bs \left( \rho \otimes P^{\downarrow}_K \right) \bsd \right],
\label{POn}
\end{equation}
where $P^{\downarrow}_K = \sum_{k=0}^K \proj{k}$. Note that $\mathcal{B}^{[K]}_{\eta}$ is not trace-preserving here since  $P^{\downarrow}_K $ is not normalized.
We need to show that $\mathcal{B}^{[K]}_{\eta} \big[ \proj{i} \big]  \succ_\mathrm{F} \mathcal{B}^{[K]}_{\eta} \big[ \proj{i+1} \big], \forall i\ge 0$, or
\begin{equation}
\rm{Tr}\left[ P_n \left( \mathcal{B}^{[K]}_{\eta} \big[ \proj{i} \big] - \mathcal{B}^{[K]}_{\eta} \big[ \proj{i+1} \big] \right) \right] \ge 0, \quad \forall i\ge 0, \forall n\ge 0.
\end{equation}
In \cite{RecBSTMS}, it was shown that if the environment is in a single Fock state $\ket{k}$, the action of the corresponding channel on a Fock state $\ket{i}$ can be written as
%  \mathcal{F}^{[k]}_{\eta}[\proj{i}] =  
\begin{equation}
 \mathrm{Tr}_E \left[ \bs \left( \proj{i} \otimes \proj{k} \right) \bsd \right] = \sum_{m=0}^{i+k} B^{(i,k)}_m \proj{m}.
\end{equation}
where the coefficients $B^{(i,k)}_m$ obey the recurrence relation
\begin{equation}
B^{(i,k)}_m = \eta B^{(i-1,k)}_{m-1} + (1-\eta ) B^{(i-1,k)}_{m} + \eta B^{(i,k-1)}_{m}  + (1-\eta) B^{(i,k-1)}_{m-1} - B^{(i-1,k-1)}_{m-1},
\label{recBS1}
\end{equation}
when $i \ge 0$, $k \ge 0$ and $0 \le m \le i+k$. Whenever one of the indices $i,k,m$ is equal to zero in the left-hand side of Eq.~(\ref{recBS1}), the coefficients with negative indices have to be removed on its right-hand side except if all indices are equal zero, in which case the "initial condition" is $B_0^{(0,0)} = 1$. Using these notations, we need to prove that
\begin{equation}
\Delta_n^{(i,K)} = \sum_{k=0}^K \sum_{m=0}^n \left[ B^{(i,k)}_m - B^{(i+1,k)}_{m} \right] \ge 0, \quad \forall i\ge 0, \forall n\ge 0, n\le i+k.
\label{eqProofLemPOLFM}
\end{equation}
Using the recurrence relation (\ref{recBS1}), we have that
\begin{eqnarray}
\hspace{-1cm} \Delta_n^{(i,K)} &=&  \sum_{k=0}^K \sum_{m=0}^n \left[ B^{(i,k)}_m - (1-\eta) B^{(i,k)}_m \right]  \nonumber \\
&& \quad - \sum_{k=0}^K \sum_{m=0}^n \left[ \eta B^{(i,k)}_{m-1} + (1-\eta) B^{(i+1,k-1)}_{m-1} + \eta B^{(i+1,k-1)}_{m} - B^{(i,k-1)}_{m-1} \right] \nonumber \\
&=& \eta \sum_{k=0}^K \sum_{m=0}^n \left( B^{(i,k)}_m - B^{(i,k)}_{m-1} \right) - \eta \sum_{k=0}^K \sum_{m=0}^n \left( B^{(i+1,k-1)}_{m} - B^{(i+1,k-1)}_{m-1}   \right) \nonumber \\
&& \quad  + \sum_{k=0}^K \sum_{m=0}^n \left( B^{(i,k-1)}_{m-1} - B^{(i+1,k-1)}_{m-1} \right) \nonumber \\
&=&  \eta \sum_{k=0}^K B^{(i,k)}_{n} - \eta \sum_{k=0}^K B^{(i+1,k-1)}_{n} + \sum_{k=0}^{K-1} \sum_{m=0}^{n-1} \left( B^{(i,k)}_{m} - B^{(i+1,k)}_{m} \right)  \nonumber \\
&=& \eta \sum_{k=0}^{K-1} B^{(i,k)}_{n} + \eta B^{(i,K)}_{n} - \eta \sum_{k=0}^{K-1} B^{(i+1,k)}_{n}  \nonumber \\
&& \quad  + \eta \sum_{k=0}^{K-1} \sum_{m=0}^{n-1} \left( B^{(i,k)}_{m} - B^{(i+1,k)}_{m} \right) + (1-\eta) \Delta_{n-1}^{(i,K-1)}   \nonumber \\
&=& \eta B^{(i,K)}_{n} + \eta \Delta_{n}^{(i,K-1)} + (1-\eta) \Delta_{n-1}^{(i,K-1)}
\end{eqnarray}
For $K=0$, we know that $\Delta_n^{(i,0)} \ge 0, \forall i\ge 0, \forall n\ge 0$ since it $\mathcal{B}^{[0]}_{\eta}$ corresponds to a Gaussian pure-loss channel \cite{Michael2016}. We are then able to prove Eq.~(\ref{eqProofLemPOLFM}) by using a recursion on $K$, since $B^{(i,k)}_{n} \ge 0, \forall i\ge 0, \forall n\ge 0, \forall k \ge 0$. This implies that 
\begin{equation}
\mathcal{B}^{[K]}_{\eta} \big[ \proj{i} \big]  \succ_\mathrm{F} \mathcal{B}^{[K]}_{\eta} \big[ \proj{i+1} \big], \quad  \forall i\ge 0.
\end{equation} Now, since any passive state can be written as a convex sum over $K$ of (normalised) projectors $P^{\downarrow}_K$, the channels $\mathcal{B}^{\downarrow}_{\eta}$ can also be written as a convex combination of channels $\mathcal{B}^{[K]}_{\eta} $, hence we get the same Fock-majorization relation for channels $\mathcal{B}^{\downarrow}_{\eta}$, which concludes the proof of Theorem \ref{lemPOLFM}. $\Box$

\noindent Using Theorems \ref{theoMichael2016_2} and  \ref{lemPOLFM}, we obtain the following Corollary.
\begin{corollary}
Passive-environment bosonic channels $\mathcal{B}^{\downarrow}_{\eta}$ are Fock-majorization preserving, that is, for all states $\rho$ and $\sigma$,
\begin{equation}
\mathrm{if~}   \rho \succ_{\rm{F}} \sigma, \mathrm{~~~then~}  \mathcal{B}^{\downarrow}_{\eta}[\rho] \succ_{\rm{F}}  \mathcal{B}^{\downarrow}_{\eta}[\sigma]
\end{equation}
\label{corPOFM}
\end{corollary}

%Here, we show that the passive channels we defined are Fock-majorization preserving, by first proving the following Lemma.

Just like Gaussian channels, passive-environment bosonic  channels do not  preserve regular majorization, that is,  if $\rho \succ \sigma$, then we cannot conclude that $\mathcal{B}^{\downarrow}_{\eta}[\rho] \succ \mathcal{B}^{\downarrow}_{\eta}[\sigma]$. Counter-examples can be easily found. However, one can prove that passive-environment bosonic channels become majorization preserving when restricting to the set of passive states. Because of the equivalence between majorization and Fock-majorization for this set, we simply need to verify that passive states remain passive after evolving through the channel. This is the content of the following Theorem.
\begin{theo} 
Passive-environment bosonic channels $\mathcal{B}^{\downarrow}_{\eta}$ are passive preserving, that is 
\begin{equation}
\mathrm{if~} \rho^{\downarrow} \mathrm{~is~passive,~~~then~} \mathcal{B}^{\downarrow}_{\eta}[\rho^{\downarrow}] \mathrm{~is~also~passive.}
\end{equation}
\label{lemPOPP}
\end{theo}
\textit{Proof.} We begin by showing that this Theorem  is true for any passive channel $\mathcal{B}^{[K]}_{\eta}$, but when the input is the (unnormalized) projector $P^{\downarrow}_I$. We need to prove that
\begin{equation}
\rm{Tr}\left[ \left( \proj{n} -\proj{n+1} \right) \mathcal{B}^{[K]}_{\eta} \big[ P^{\downarrow}_I \big] \right] \ge 0, \quad \forall I\ge 0, \forall n\ge 0,
\end{equation}
or,
\begin{equation}
\Gamma_n^{(I,K)} =  \sum_{i=0}^I \sum_{k=0}^K \left( B^{(i,k)}_{n} - B^{(i,k)}_{n+1} \right)  \ge 0, \quad \forall I\ge 0, \forall n\ge 0.
\label{eqProofLemPOPP}
\end{equation}
Using the recurrence relation (\ref{recBS1}), we have that
\begin{eqnarray}
\hspace{-1cm} \Gamma_n^{(I,K)} &=  \sum_{i=0}^I \sum_{k=0}^K \left( B^{(i,k)}_{n} - \eta B^{(i-1,k)}_{n} - (1-\eta ) B^{(i-1,k)}_{n+1} - \eta B^{(i,k-1)}_{n+1}    \right.  \nonumber \\
& \quad\quad   \left.  - (1-\eta) B^{(i,k-1)}_{n} + B^{(i-1,k-1)}_{n} \right)   \nonumber \\
& = \eta \sum_{i=0}^I \sum_{k=0}^K \left( B^{(i,k)}_{n} - B^{(i-1,k)}_{n} \right) + (1-\eta) \sum_{i=0}^I \sum_{k=0}^K \left( B^{(i,k)}_{n} - B^{(i,k-1)}_{n} \right) \nonumber \\
& \quad \quad - (1-\eta ) \sum_{i=0}^I \sum_{k=0}^K B^{(i-1,k)}_{n+1} - \eta \sum_{i=0}^I \sum_{k=0}^K B^{(i,k-1)}_{n+1} + \sum_{i=0}^I \sum_{k=0}^K B^{(i-1,k-1)}_{n}   \nonumber \\
&= \eta \sum_{k=0}^K B^{(i,k)}_{n} + (1-\eta) \sum_{i=0}^I B^{(i,k)}_{n} - (1-\eta ) \sum_{i=0}^{I-1} \sum_{k=0}^K B^{(i,k)}_{n+1}   \nonumber \\
& \quad\quad  - \eta \sum_{i=0}^I \sum_{k=0}^{K-1} B^{(i,k)}_{n+1} + \sum_{i=0}^{I-1} \sum_{k=0}^{K-1} B^{(i,k)}_{n}   \nonumber \\
&= B^{(i,k)}_{n} - (1-\eta ) \sum_{i=0}^{I-1} \sum_{k=0}^K B^{(i,k)}_{n+1} - \eta \sum_{i=0}^I \sum_{k=0}^{K-1} B^{(i,k)}_{n+1} + \eta \sum_{i=0}^{I} \sum_{k=0}^{K-1} B^{(i,k)}_{n}  \nonumber \\
& \quad\quad + (1-\eta) \sum_{i=0}^{I-1} \sum_{k=0}^{K} B^{(i,k)}_{n}   \nonumber \\
&= B^{(i,k)}_{n} + \eta \sum_{i=0}^{I} \sum_{k=0}^{K-1} \left( B^{(i,k)}_{n} - B^{(i,k)}_{n+1} \right) + (1-\eta) \sum_{i=0}^{I-1} \sum_{k=0}^{K} \left( B^{(i,k)}_{n} - B^{(i,k)}_{n+1} \right)    \nonumber \\
&= B^{(i,k)}_{n} + \eta \Gamma_n^{(I,K-1)} + (1-\eta) \Gamma_n^{(I-1,K)}
\end{eqnarray}
We know that $\Gamma_n^{(I,0)} \ge 0, \forall I\ge 0, \forall n\ge 0$, since it corresponds to a Gaussian pure-loss channel, and was proven in \cite{Michael2016}. We also know, because of the symmetry of the beam splitter, that $\Gamma_n^{(0,K)} \ge 0, \forall K\ge 0, \forall n\ge 0$. We are then able to prove (\ref{eqProofLemPOPP}) by using a recursive argument on both $I$ and $K$, since $B^{(i,k)}_{n} \ge 0, \forall I\ge 0, \forall K \ge 0, \forall n\ge 0$. This shows that $\mathcal{B}^{[K]}_{\eta}[P^{\downarrow}_I]$ is passive. As before, we conclude the proof by using the fact that any passive state can be written as a convex sum of (normalised) projectors $P^{\downarrow}_l$. $\Box$

Using Corollary \ref{corPOFM} and Theorem \ref{lemPOPP}, we are now able to state the following.
\begin{corollary}
Passive-environment bosonic channels $\mathcal{B}^{\downarrow}_{\eta}$ are majorization-preserving over the set of passive states, 
that is, for any two passive states $\rho^{\downarrow}$ and $\sigma^{\downarrow}$,
\begin{equation}
\mathrm{if~}   \rho^{\downarrow} \succ \sigma^{\downarrow}, \mathrm{~~~then~}  \mathcal{B}^{\downarrow}_{\eta}[\rho^{\downarrow}] \succ  \mathcal{B}^{\downarrow}_{\eta}[\sigma^{\downarrow}]
\end{equation}
\label{corPOM}
\end{corollary}

\section{Passive-environment channels with an active Gaussian unitary}

For completeness, we now show that all the results of Section~4 extend to the passive-environnement channels obtained with an active Gaussian unitary, namely $\mathcal{A}^{\downarrow}_{G}$. We will recourse to the following theorem.
\begin{theo}[\cite{Michael2016}]
A channel $\Phi$ satisfying the condition $\bra{n} \, \Phi[\ket{i}\bra{i}] \, \ket{m}=0$,  $\forall n \ne m$, $\forall i$ is passive preserving if and only if its adjoint channel $\phi^\dagger$ obeys the ladder of Fock-majorization relations
\begin{equation}
\Phi^\dagger \big[ \proj{i} \big]  \succ_\mathrm{F} \Phi^\dagger \big[ \proj{i+1} \big], \quad  \forall i\ge 0.   \label{ladder2}
\end{equation}
\label{2ndtheorem}
\end{theo}
Again, compared to the statement of Theorem \ref{2ndtheorem} in \cite{Michael2016}, we need to add the condition that the diagonal elements at the input of the channel do not yield non-diagonal elements at the output of the channel. This condition is fullfiled for Gaussian-dilatable channels with a passive environment, see \ref{appNonDiag}. In order to exploit Theorem \ref{2ndtheorem}, the last thing that remains to be done is to prove the duality between channels $\mathcal{B}^{\downarrow}_{\eta}$ and $\mathcal{A}^{\downarrow}_{G}$. This is the content of our last theorem.
\begin{theo} 
A channel $\mathcal{C}_{\eta}^{\mathrm{BS}}$ whose Stinespring dilation is based on a beam-splitter of transmittance $\eta$, \textit{i.e.}
\begin{equation}
	\mathcal{C}_{\eta}^{\mathrm{BS}}[\bullet] = \mathrm{Tr}_{\mathrm{E}} \left[ \bs \left( \bullet \otimes \sigma_{\mathrm{E}}^{(1)} \right) \bsd \right],
\end{equation}
is dual to a channel $\mathcal{C}_{\lambda}^{\mathrm{TMS}}$ having a two-mode squeezer of parameter $\lambda=\tanh^2(r)$ in its Stinespring dilation ($r$ being the squeezing parameter), \textit{i.e.}
\begin{equation}
	\mathcal{C}_{\lambda}^{\mathrm{TMS}}[\bullet] = \mathrm{Tr}_{\mathrm{E}} \left[ \tms \left( \bullet \otimes \sigma_{\mathrm{E}}^{(2)} \right) \tmsd \right],
\end{equation}
with $\lambda=1-\eta$ and $\sigma_{\mathrm{E}}^{(2)} = \left(\sigma_{\mathrm{E}}^{(1)}\right)^{\mathrm{T}} \!\!\!$, where $T$ denotes matrix transposition in the Fock basis.
\label{theorem-duality}
\end{theo} 
\textit{Proof.}  We start with two states expressed in the Fock basis as
\begin{equation}
	\rho = \sum_{i,j} \rho_{i,j} \kb{i}{j}, \quad \mathrm{and} \quad \gamma = \sum_{n,m} \gamma_{n,m} \kb{n}{m},
\end{equation}
and compute the object
\begin{equation}
	\mathrm{Tr}\left[\gamma \ \mathcal{C}_{\eta}^{\mathrm{BS}}[\rho]\right] = \sum_{n,m} \gamma_{n,m} \sum_{i,j} \rho_{i,j} \bra{m} \mathcal{C}_{\eta}^{\mathrm{BS}}\left[\kb{i}{j}\right] \ket{n}.
\end{equation}
If we consider a general environment $\sigma_{\mathrm{E}}^{(1)} = \sum_{k,l} \sigma^{(1)}_{k,l} \kb{k}{l}$, we get
\begin{equation*}
	\mathrm{Tr}\left[\gamma \ \mathcal{C}_{\eta}^{\mathrm{BS}}[\rho]\right] = \sum_{n,m} \gamma_{n,m} \sum_{i,j} \rho_{i,j} \sum_{k,l} \sigma^{(1)}_{k,l} \sum_e \bra{m,e} \bs \kb{i,k}{j,l} \bsd \ket{n,e}.
\end{equation*}
It was shown in \cite{RecBSTMS} that, under partial time reversal, a beam splitter is turned into a two-mode squeezer, or, more precisely, their respective transition amplitudes in the Fock basis are related through
\begin{equation}
	\bra{m,e} \bs \ket{i,k} = \frac{1}{\sqrt{\eta}} \bra{m,k} \tms \ket{i,e},
	\label{eqPTR}
\end{equation}
where $\lambda = 1-\eta$. Notice that the ket and bra of the second mode have been swapped in Equation (\ref{eqPTR}). This property leads to
\begin{eqnarray*}
\fl	\mathrm{Tr}\left[\gamma \ \mathcal{C}_{\eta}^{\mathrm{BS}}[\rho]\right] = \frac{1}{\eta} \sum_{n,m} \gamma_{n,m} \sum_{i,j} \rho_{i,j} \sum_{k,l} \sigma^{(1)}_{k,l} \sum_e \bra{m,k} \tms \kb{i,e}{j,e} \tmsd \ket{n,l} & \\
	= \frac{1}{\eta} \sum_{n,m} \gamma_{n,m} \sum_{i,j} \rho_{i,j} \sum_{k,l} \sigma^{(1)}_{k,l} \sum_e \bra{j,e} \tmsd \kb{n,l}{m,k} \tms \ket{i,e} & .
\end{eqnarray*}
Now, one understands that the two-mode squeezer $\tms$ has the same effect as $\tmsd$ in a channel of the form of $\mathcal{C}_{\lambda}^{\mathrm{TMS}}$ since there is no difference at the level of probabilities. As a consequence,
\begin{equation*}
	\mathrm{Tr}\left[\gamma \ \mathcal{C}_{\eta}^{\mathrm{BS}}[\rho]\right] = \frac{1}{\eta} \sum_{n,m} \gamma_{n,m} \sum_{i,j} \rho_{i,j} \bra{m} \mathcal{C}_{1-\eta}^{\mathrm{TMS}}\left[\kb{i}{j}\right] \ket{n} = \mathrm{Tr}\left[\rho \ \mathcal{C}_{1-\eta}^{\mathrm{TMS}}[\gamma]\right] ,
\end{equation*}
where the environment $\sigma_{\mathrm{E}}^{(2)}$ characterizing the channel $\mathcal{C}_{1-\eta}^{\mathrm{TMS}}$ is related to the environment  $\sigma_{\mathrm{E}}^{(1)}$ of $\mathcal{C}_{\eta}^{\mathrm{BS}}$ through $\sigma_{\mathrm{E}}^{(2)} = \left(\sigma_{\mathrm{E}}^{(1)}\right)^{\mathrm{T}} \!\!\!$. In other words, the adjoint of channel $\mathcal{C}_{\eta}^{\mathrm{BS}}$ verifies
\begin{equation}
	\left(\mathcal{C}_{\eta}^{\mathrm{BS}}\right)^{\dagger} = \frac{1}{\eta} \, \mathcal{C}_{1-\eta}^{\mathrm{TMS}},
\end{equation}
where one should transpose the environment in the Fock basis.
$\Box$

In a two-mode squeezer, the parameter $\lambda$ is related to the parametric gain $G$ via $\lambda = (G-1)/G$, so that the relation $\lambda = 1-\eta$ translates into  $G = 1/\eta$. Thus, in the special case of passive-environment channels (for which the transpose of the environment state remains unchanged) the adjoint channel of 
$\mathcal{B}^{\downarrow}_{\eta}$  is $\frac{1}{\eta}\mathcal{A}^{\downarrow}_{1/\eta}$, in full analogy with the situation for Gaussian channels. This allows us to state the following corollaries.
\begin{corollary}
Passive-environment bosonic channels $\mathcal{A}^{\downarrow}_{G}$ are Fock-majorization preserving, that is, for all states $\rho$ and $\sigma$,
\begin{equation}
\mathrm{if~}   \rho \succ_{\rm{F}} \sigma, \mathrm{~~~then~}  \mathcal{A}^{\downarrow}_{G}[\rho] \succ_{\rm{F}}  \mathcal{A}^{\downarrow}_{G}[\sigma]
\end{equation}
\label{corol3}
\end{corollary}
Indeed, since $\mathcal{B}^{\downarrow}_{\eta}$ is  passive-preserving (Theorem \ref{lemPOPP}), the duality property of passive-environment channels (Theorem \ref{theorem-duality}) combined with Theorem \ref{2ndtheorem} implies that $\mathcal{A}^{\downarrow}_{G}$ satisfies the ladder of Fock-majorization relations
$\mathcal{A}^{\downarrow}_{G} \big[ \proj{i} \big]  \succ_\mathrm{F} \mathcal{A}^{\downarrow}_{G} \big[ \proj{i+1} \big]$,   $\forall i\ge 0$, hence it is Fock-majorization preserving as a consequence of Theorem \ref{theoMichael2016_2}.

\begin{corollary} 
Passive-environment bosonic channels $\mathcal{A}^{\downarrow}_{G}$ are passive preserving, that is 
\begin{equation}
\mathrm{if~} \rho^{\downarrow} \mathrm{~is~passive,~~~then~} \mathcal{A}^{\downarrow}_{G}[\rho^{\downarrow}] \mathrm{~is~also~passive.}
\end{equation}
\end{corollary}
Indeed, since $\mathcal{B}^{\downarrow}_{\eta}$ satisfies the ladder of Fock-majorization relations (\ref{EQlemPOLFM}) (Theorem \ref{lemPOLFM}), 
the duality property of passive-environment channels (Theorem \ref{theorem-duality}) combined with Theorem \ref{2ndtheorem} 
implies that $\mathcal{A}^{\downarrow}_{G}$ is passive-preserving.

Finally, using these two corollaries and the equivalence between majorization and Fock-majorization for the set of passive states, we obtain the following.
\begin{corollary}
Passive-environment channels $\mathcal{A}^{\downarrow}_{G}$ are majorization-preserving over the set of passive states,
that is, for any two passive states $\rho^{\downarrow}$ and $\sigma^{\downarrow}$,
\begin{equation}
\mathrm{if~}   \rho^{\downarrow} \succ \sigma^{\downarrow}, \mathrm{~~~then~}  \mathcal{A}^{\downarrow}_{G}[\rho^{\downarrow}] \succ  \mathcal{A}^{\downarrow}_{G}[\sigma^{\downarrow}]
\end{equation}
\label{corol5}
\end{corollary}
%This extends Corollary \ref{corPOM} to passive-environment channels with an active Gaussian unitary.

\section{Conclusion}

In summary, we have shown that all bosonic quantum channels whose Stinespring dilation involves a Gaussian unitary (either a beam splitter or a two-mode squeezer) and a passive environment (from which no energy can be extracted by acting with a unitary) exhibit a series of properties regarding how the order or disorder (measured via majorization) is transfered across the channel. Our central result is that any such channel preserves the Fock-majorization relation (see Corollary \ref{corPOFM} for channel $\mathcal{B}^{\downarrow}_{\eta}$ and Corollary \ref{corol3} for $\mathcal{A}^{\downarrow}_{G}$). Moreover, as a consequence of being passive-preserving, all these passive-environment channels also preserve the regular majorization relation over the set of passive states (see Corollary \ref{corPOM} for $\mathcal{B}^{\downarrow}_{\eta}$ and Corollary \ref{corol5} for $\mathcal{A}^{\downarrow}_{G}$).

These results heavily rely on the Fock-majorization relation for bosonic systems. Because of its connection with energy, Fock-majorization can be viewed as the fundamental mathematical relation that is conserved when quantum states evolve through passive-environment channels, which allows one to relate the concepts of disorder and energy (when dealing with passive states, the concepts of majorization and Fock-majorization become equivalent). Our paper can thus be read in the context of quantum thermodynamics, where we define the class of passive-environment operations that encompass -- but go beyond -- thermal operations and characterise the properties of such operations in terms of majorization theory. These results will hopefully contribute to connect the area of continuous-variable bosonic channels together with quantum thermodynamics.

More generally, the notions of passive-environment channel and Fock-majorization relation are independent of the specific nature of the considered system, so we anticipate that our results can be extended to other quantum systems (beyond a bosonic mode) and arbitrary Hamiltonians (beyond the Harmonic oscillator). The energy-majorization relation between two states (based on comparing the diagonal elements in the energy eigenbasis) should then be conserved along the thermodynamical operation resulting from the energy-conserving coupling of the system with a passive environment (i.e., an environment state having the minimum energy compatible with its eigenspectrum).

%This question is of strong interest since majorization is a powerful theory with many applications in quantum information and quantum thermodynamics among others. 

%%%%%%%%%%%%%%%%%%%%%%%%%%%%%%%%%%%%%%%%% Acknowledgments

\ack{We thank Ognyan Oreshkov for useful discussions. This work was supported by the F.R.S.-FNRS under Project No. T.0224.18. M.G.J. also acknowledges support from the FRIA foundation.}

%%%%%%%%%%%%%%%%%%%%%%%%%%%%%%%%%%%%%%%%% Appendix

\appendix

\section{Extension of Fock-majorization preservation to non-diagonal states in the case of passive channels \label{appNonDiag}}

Here, we prove that Theorem \ref{theoMichael2016_2} can be extended to states which are non-diagonal in the Fock-basis when $\Phi \equiv \mathcal{B}^{\downarrow}_{\eta}$. We do this by showing that if $\rho$ is diagonal in the Fock basis, $\mathcal{B}^{\downarrow}_{\eta}[\rho]$ is also diagonal in the Fock basis, while if $\rho$ is non-diagonal in the Fock basis, its non-diagonal elements do not contribute to the diagonal elements of $\mathcal{B}^{\downarrow}_{\eta}[\rho]$. It can be shown that \cite{RecBSTMS}
\begin{equation}
\bs \ket{i,k} = \sum_{n=0}^{i+k} \xi_n^{(i,k)} \ket{n,i+k-n},
\end{equation}
where $\xi_n^{(i,k)} \in \mathbb{C}$. If we define our passive channel as in Eq. (\ref{PO}), we have
\begin{eqnarray}
\mathcal{B}^{\downarrow}_{\eta}[\kb{i}{j}] & = \sum_k \lambda^{\downarrow}_k \mathrm{Tr}_E \left[ \bs \left( \kb{i}{j} \otimes \proj{k} \right) \bsd \right] \nonumber \\
& = \sum_k \lambda^{\downarrow}_k \sum_l \bra{l}_E \left( \sum_{n=0}^{i+k} \xi_n^{(i,k)} \ket{n,i+k-n} \right) \nonumber \\
& \quad \quad \quad \quad \quad \quad \quad \quad \times \left( \sum_{m=0}^{j+k} \xi_m^{(j,k)*} \bra{m,j+k-m} \right) \ket{l}_E \nonumber \\
& = \sum_k \lambda^{\downarrow}_k \sum_n \xi_n^{(i,k)} \xi_{n+j-i}^{(j,k)*} \kb{n}{n+j-i} \label{eqappNonDiag1}
\end{eqnarray}
We end up with
\begin{equation}
\mathcal{B}^{\downarrow}_{\eta}[\kb{i}{i}] = \sum_k \lambda^{\downarrow}_k \sum_n |\xi_n^{(i,k)}|^2 \kb{n}{n},
\end{equation}
which means that if $\rho$ is diagonal in the Fock basis, $\mathcal{B}^{\downarrow}_{\eta}[\rho]$ is also diagonal in the Fock basis. Furthermore, Eq. (\ref{eqappNonDiag1}) tells us that if $\rho$ is non-diagonal in the Fock basis, its non-diagonal elements do not contribute to the diagonal elements of $\mathcal{B}^{\downarrow}_{\eta}[\rho]$.

%%%%%%%%%%%%%%%%%%%%%%%%%%%%%%%%%%%%%%%%% Bibliography


\section*{References}

\begin{thebibliography}{99}

\bibitem{qthermo1} J. Åberg, Nat. Commun. \textbf{4}, 1925 (2013).

\bibitem{qthermo2} M. Horodecki and J. Oppenheim, Nat. Commun. \textbf{4}, 2059 (2013).

\bibitem{qthermo3} P. Skrzypczyk, A. J. Short, and S. Popescu, Nat. Commun. \textbf{5}, 4185 (2014).

\bibitem{qthermo4} G. Gour, M. P. M\"uller, V. Narasimhachar, R. W. Spekkens, and N. Y. Halpern, Phys. Rep. \textbf{583}, 1 (2015).

\bibitem{qthermo5} F. Brandão, M. Horodecki, N. Ng, J. Oppenheim, and S. Wehner, Proc. Nat. Acad. Sci. \textbf{112}, 3275 (2015). 



\bibitem{Ineq} A. W. Marshall and I. Olkin, \textit{Inequalities: Theory of Majorization and its Applications} (Academic Press, New York, 1979).%

\bibitem{NielsenLOCC} M. A. Nielsen, Phys. Rev. Lett. \textbf{83,} 436 (1999).%

\bibitem{Michael2015} M. G. Jabbour, R. Garc\'ia-Patr\'on, and N. J. Cerf, Phys. Rev. A \textbf{91}, 012316 (2015).%

\bibitem{Michael2016} M. G. Jabbour, R. Garc\'ia-Patr\'on, and N. J. Cerf, New J. Phys. \textbf{18}, 073047 (2016).%

\bibitem{RecBSTMS} M. G. Jabbour and N. J. Cerf, arXiv:1803.10734 [quant-ph].

\bibitem{Passive1} W. Pusz and S. L. Woronowicz, Commun. Math. Phys. \textbf{58}, 273 (1978).%

\bibitem{Passive2} A Lenard, J. Stat. Phys. \textbf{19}, 575 (1978).%

\bibitem{Thermo} J. Goold, M. Huber, A. Riera, L. del Rio, and P. Skrzypczyk, J. Phys. A: Math. Theor. \textbf{49}, 143001 (2016).%

\bibitem{MajNO} M. Horodecki, P. Horodecki, and J. Oppenheim, Phys. Rev. A \textbf{67}, 062104 (2003).%

\bibitem{Weedbrook2011} C. Weedbrook, S. Pirandola, R. Garc\'{\i}a-Patr\'{o}n, T. Ralph, N. J. Cerf, J. H. Shapiro, and S. Lloyd, Rev. Mod. Phys. \textbf{84,} 621 (2012).%

\bibitem{qt1} V. Narasimhachar and G. Gour, Nature Comm. \textbf{6}, 7689 (2015).%


\end{thebibliography}
\end{document}